\documentclass[12pt,preprint]{aastex}

\newcommand{\mydeg}{\mbox{$^{\circ}$}}

\newcommand{\kms}{\mbox{km s$^{-1}$}}

\newcommand{\ha}{\mbox{H$\alpha$}}
\newcommand{\hb}{\mbox{H$\beta$ }}

\newcommand{\lam}{\mbox{$\lambda$}}
\newcommand{\mum}{\mbox{$\mu$m}}

\newcommand{\msun}{\mbox{M$_{\odot}$}}

\shorttitle{AGN Feedback in a High-z Radio Galaxy}

\begin{document}

\title{Extreme gas kinematics in the z=2.2 powerful radio galaxy
MRC1138-262: Evidence for efficient AGN feedback in the early
Universe?\altaffilmark{1}}

\author{N.~P.~H. Nesvadba,\altaffilmark{2}
  M.~D.~Lehnert,\altaffilmark{2}
  F.~Eisenhauer,\altaffilmark{2} A.~Gilbert,\altaffilmark{2,3}
  M.~Tecza,\altaffilmark{4} \& R.~Abuter\altaffilmark{2,5}}

\altaffiltext{1}{Based on observations collected at the European Southern
Observatory, Very Large Telescope Array, Cerro Paranal;
program numbers, 70.B-0545, 70.A-0229, \& 076.A-0684}

\altaffiltext{2}{Max-Planck-Institut f\"ur extraterrestrische Physik,
Giessenbachstra\ss e, 85748 Garching bei M\"unchen, Germany}

\altaffiltext{3}{Current Address: Institute of Geophysics and Planetary
Physics, Lawrence Livermore National Laboratory, 7000 East Avenue, L413,
Livermore, CA 94550}

\altaffiltext{4}{University of Oxford, Subdepartment of Astrophysics,
Denys Wilkinson Building, Keble Road, Oxfordshire, Oxford OX1 3RH,
United Kingdom}

\altaffiltext{5}{Current Address: European Southern Observatory,
Karl-Schwarzschild-Strasse 2, Garching D-85748, Germany}


\begin{abstract}
To explain the properties of the most massive low-redshift galaxies
and the shape of their mass function, recent models of galaxy evolution
include strong AGN feedback to complement starburst-driven feedback in
massive galaxies. Using the near-infrared integral-field spectrograph
SPIFFI on the VLT, we searched for direct evidence for such a feedback
in the optical emission line gas around the z$=$2.16 powerful radio
galaxy MRC1138-262, likely a massive galaxy in formation.  The kpc-scale
kinematics, with FWHMs and relative velocities $\lesssim 2400$ \kms\
and nearly spherical spatial distribution, do not resemble large-scale
gravitational motion or starburst-driven winds. Order-of-magnitude
timescale and energy arguments favor the AGN as the only plausible
candidate to accelerate the gas, with a total energy injection of
$\sim few \times 10^{60}$ ergs or more, necessary to power the
outflow, and relatively efficient coupling between radio jet and ISM.
Observed outflow properties are in gross agreement with the models,
and suggest that AGN winds might have a similar, or perhaps larger,
cosmological significance than starburst-driven winds, if MRC1138-262
is indeed archetypal. Moreover, the outflow has the potential to
remove significant gas fractions ($\lesssim 50$\%) from a ${>\cal
L}^{*}$ galaxy within a few 10 to 100 Myrs, fast enough to preserve the
observed [$\alpha$/Fe] overabundance in massive galaxies at low redshift.
Using simple arguments, it appears that feedback like that observed in
MRC1138-262 may have sufficient energy to inhibit material from infalling
into the dark matter halo and thus regulate galaxy growth as required
in some recent models of hierarchical structure formation.  \end{abstract}


\section{Introduction}
Models of structure formation and galaxy evolution have reached a state
where the impact of large-scale baryonic feedback on galaxy evolution
can no longer be neglected. However, the physics of such processes
are complex, and as yet not very well understood, especially at high
redshift, where the most massive galaxies formed most of their stars.
Therefore, simulations 
of the ensemble of galaxies typically incorporate these processes,
which originate on scales beyond the resolution of typical large scale
simulations, as simple relationships and parameterizations, which are in
turn tuned to approximately match observed galaxy properties. In spite
of these limitations, baryonic feedback plays a crucial role for our
understanding of galaxy evolution, perhaps most dramatically at high
redshift \citep[e.g.,][]{croton06}

The impact of feedback related to intense star-formation is now
observationally well-established at low and high redshift
\citep[e.g.,][]{heckman03, lehnert96a}, and simulations improved
considerably when including it. But significant and likely fundamental
discrepancies remain, e.g., when comparing the {\it observed} mass
function of galaxies with the {\it predicted} mass function of dark matter
halos at the upper end of the mass function \citep{benson03}. From
low-redshift studies, starburst-driven ``superwinds'' are known
to be efficient in driving out gas only from the comparably shallow
gravitational potentials of low mass galaxies \citep[e.g.,][]{martin99,
lehnert99, heckman00}. However, these winds are mainly energy driven,
therefore susceptible to radiative losses, and they probably lack the
power, efficiency, and velocities necessary to remove significant gas
fractions from the deep potential wells of the most massive galaxies
and their dark-matter halos. As a consequence, they cannot explain the
overabundance particularly of massive dark-matter halos compared to the
co-moving galaxy density.

Therefore another powerful feedback mechanism is recently
gaining in popularity among researchers modeling and simulating
the ensemble properties of galaxies, which are vigorous outflows
driven by powerful active galactic nuclei at high redshift \citep[AGN;
e.g.,][]{silk98,springel05}.  Energy- and momentum-driven AGN winds may
reach very high energy output through radiation pressure and particle
ejection.  The most powerful AGN may eject enough energy and momentum
to unbind significant fractions of the total interstellar medium (ISM)
of the host galaxy. As they slow and expand, such winds are able to
efficiently heat gas in the inter-cluster medium (ICM) and even the
intergalactic medium \citep[IGM; e.g.,][]{nath02}. However, due to the
clumpy nature of the ISM in galaxies, rather small working surface of
the jet, and AGN producing ``light jets'', they are not generally thought
to couple strongly with the ambient ISM in their hosts \citep{begelman89}.

The observed characteristics of massive galaxies support the hypothesis
that strong AGN winds at high redshift may have a significant impact
on galaxy evolution and their ensemble characteristics.  Spheroidal
and black hole masses in nearby massive systems follow the tight
M$_{BH}$-$\sigma$ relation \citep{ferrarese02, tremaine02}, which perhaps
signals co-eval growth of SMBH and the bulge self-regulated through
negative AGN feedback \citep{silk98}. The most massive galaxies,
virtually all spheroidals, are ``old, red, and dead''. They are
metal-rich and have luminosity weighted ages consistent with massive
star-formation at high-redshift and subsequent passive evolution,
perhaps indicating that the gas was effectively removed during the most
massive burst \citep[e.g.,][]{thomas99,thomas05,romano02}.  Maybe the
tightest constraint comes from the metal abundance ratios, especially
$[\alpha/Fe]$, which indicate intense star-formation truncated by strong
feedback within less than few $\times 10^8$ yrs \citep[e.g.,][]{thomas99}.
In addition, the exponential cut-off in the bright end of the galaxy
luminosity function may be a direct result of the efficiency of AGN
feedback \citep{croton06}. Furthermore, \citet{best05} find evidence
that more moderate feedback of radio-loud AGN might balance gas cooling
in the halos of the most massive galaxies at low redshift.

A promising way to identify the fingerprints of a highly efficient mode
of AGN feedback at high redshift is by observing the gas kinematics
in rapidly growing, massive galaxies at high redshift which host
AGN. Powerful high-z Radio Galaxies (HzRGs) fit these requirements:
They reside in dense environments \citep{venemans02, kurk04, best03,
stevens03}. Their magnitudes, colors, and continuum morphologies
are consistent with large stellar masses and passive evolution
at $z\lesssim 2-3$ \citep{vanbreugel98, best98}. HzRGs at $z >
2-3$ also contain large masses of molecular gas \citep[$10^{11}$
\msun; e.g.,][]{rottgering97}, dust \citep[$10^{8-9}$ \msun;
e.g.,][]{archibald01, reuland04}, and UV/optical line-emitting
gas \citep[$10^{8-9}$ \msun; e.g.,][]{vanoijk97}. At the highest
redshifts, they are also likely to be forming stars at prodigious rates
\citep[$100-1000$ \msun\ yr$^{-1}$; e.g.,][]{dey97}. Curiously, HzRGs
are known from longslit-spectroscopy to often have spatially extended
broad emission lines with FWHM$ > 1000$ \kms, indicative of extreme
kinematics \citep[e.g.,][]{mccarthy96,tadhunter91,villar99}. However,
with the slits typically aligned along the radio axis, the global role
of these kinematics could not be examined more closely, highlighting the
need for integral-field observations of HzRGs, preferably at rest-frame
optical wavelengths (i.e., in the observed near-infrared) to investigate
the overall gas dynamics.

An excellent candidate for such a study is the $z=2.16$ MRC1138-262,
with $H=18.04$ and stellar mass $M\sim 5\times 10^{11}$ \msun. VLA
interferometry reveals 2 radio jets 
with complex morphologies \citep{carilli02}. Its nebulosity extends
over $\sim 20-25$ kpc radius\footnote{Using a flat $H_0 = 70$\kms,
$\Omega_M=0.3$, $\Omega_{\Lambda}=0.7$ cosmololgy, size scales at
$z=2.16$ are 8.3 kpc arcsec$^{-1}$, the luminosity distance is 17070
Mpc, the angular size distance is 1710 Mpc. Cosmic age at that redshift
is 3.0 Gyrs.}, and has a significant surrounding over-density of \ha\
emitters which indicates it might be in the center of a forming galaxy
cluster \citep{kurk04}. Rest-frame UV spectroscopy implies star-formation
rates of $40-70$ \msun\ yr$^{-1}$.  By examining the spatially-resolved
kinematics and spectral properties of the rest-frame optical emission
line gas in MRC1138-262, we wish to investigate the influence of the
AGN on the excitation and kinematics of the emission line gas using data
obtained with the near-infrared integral-field spectrograph SPIFFI.

\section{Observations and Data Reduction}
\label{sec:datred}

We observed MRC1138-262 in the near-infrared H and K bands using the
integral-field spectrograph SPIFFI on UT1 of the VLT in April 2003. SPIFFI
uses an image slicer to dissect the 8\arcsec$\times$8\arcsec\ field of
view on the sky into 32 slices or ``slitlets'' with individual pixel
sizes in projection of 0.25\arcsec. The spectrograph and image slicer
have since become a facility instrument at UT4 as part of SINFONI. Under
good and stable weather conditions, we obtained a total of 140 min
of data in K and 75 min in H with a pixel scale of 0.25\arcsec\
and seeing of $\sim 0.4\arcsec \times 0.6\arcsec$.  Individual
exposure times were 300s in H and 600s in K, respectively. Spectral
resolution is R=${{\lam}\over{\Delta\lam}}\approx 2000$ at 1.6 \mum\
and R=${{\lam}\over{\Delta\lam}}\approx 2400$ at 2.2 \mum. One ``off''
frame at a sky position well displaced from the target was taken for each
``on'' frame in an off-on-on-off pattern.

To reduce the data, we used the IRAF \citep{tody93} standard tools
for the reduction of longslit-spectra, modified to meet the special
requirements of integral-field spectroscopy. Data are dark frame
subtracted and flat-fielded. For bad pixel correction we use the full
three-dimensional information, identifying pixels that are 5$\sigma$
deviant in the darks and flats. Rectification and
wavelength calibration are done before night sky subtraction, to account
for some spectral flexure between the frames. Curvature is measured and
removed using an arc lamp, before shifting the spectra to an absolute
(vacuum) wavelength scale with reference to the OH lines in the data.

The three-dimensional data cubes are then reconstructed, assuming that
each slitlet covers exactly 32 pixels. They are spatially aligned
by cross-correlating the collapsed cubes, and then combined, clipping
deviant pixels. Telluric correcton is applied to the combined cube. Flux
scales are obtained from standard star observations. From the light
profile of the standard star, we measure the FWHM spatial resolution
to be $0.6\arcsec\times0.4\arcsec$ in right ascension and declination,
respectively.

\section{Integrated Spectra and Broad Line Emission}
\label{sec:spectra}

\begin{figure}[htb]
\epsscale{0.7}
\plotone{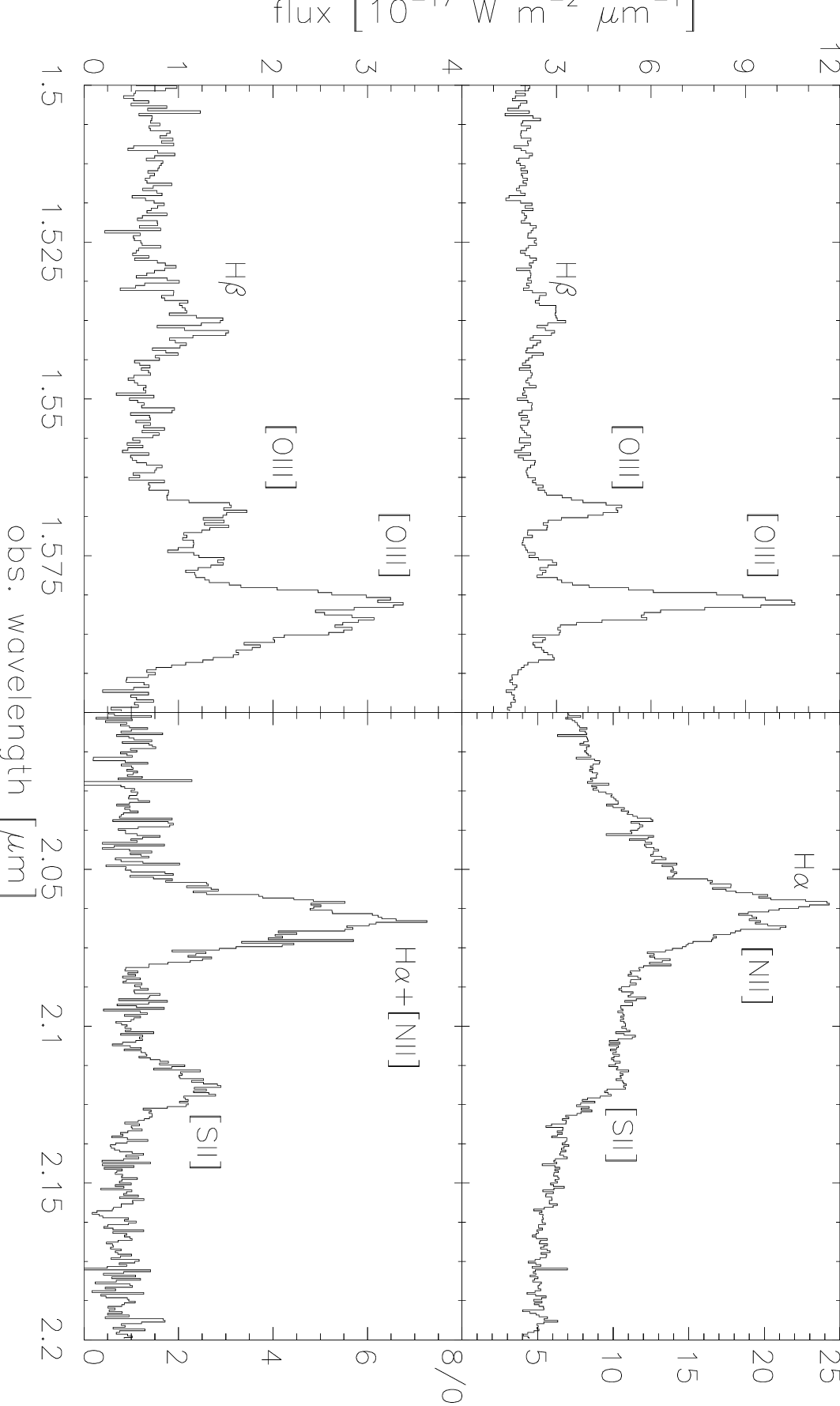}
\caption{H band (left panel) and K band (right panel) spectra of
MRC1138-262. Upper panels indicate spectra of the spatial pixel covering
the AGN, lower panels show spectra integrated over 15 kpc radius from
the nucleus, with the AGN removed (see text for details).}
\label{fig:specnuke}
\end{figure}

In Fig.~\ref{fig:specnuke} we show H and K-band spectra of
MRC1138-262. The upper panel shows the spectrum of the spatial pixel
covering the AGN, the lower panel shows the AGN removed extended emission
integrated over the central 30 kpc. Line profiles are not simple
Gaussians, but have irregular profiles indicating complex kinematics
in the extended emission line gas. The lines are generally broad,
and [NII]\lam6583 and \ha\ are blended, so that \ha\ is not a reliable
kinematic tracer. Therefore we use [OIII]\lam5007 (with FWHM$=3990$ \kms\
in the integrated spectrum) to trace the gas kinematics.  We discuss below
that the emission line kinematics does not vary strongly with the specific
emission line analyzed (\S~\ref{sec:iontempdens}), within the accuracy we
can reach given the line blending. Thus we do not expect and there is no
evidence that this choice has a strong impact on our overall conclusions.

The luminous broad \ha\ emission line is the most prominent spectral
feature in the AGN spectra, with FWHM$=14900$ \kms, and a rest-frame
equivalent width, EQW$\sim 325$\AA, indicating that the emission from the
broad line region cannot be neglected (especially in the broad band near
infrared magnitudes). Previous rest-frame UV spectroscopy of MRC1138-262
did not reveal nuclear broad lines, indicating that MRC1138-262 might
harbor an obscured quasar, similar to MRC2025-218 \citep{larkin00}. Broad
\hb\ emission in the H band is measured at much fainter flux: From the
Balmer decrement in the broad components, $A_V =8.7$, using a galactic
extinction law and $\ha/\hb\ = 2.9$. The galactic extinction law might
not be strictly adequate for broad AGN emission lines, nonetheless,
the large extinction underlines the interpretation that MRC1138-262
harbors an obscured quasar.

To correct the K-band data for the AGN contribution, we fit and remove
the nuclear spectrum from all other spatial pixels. This may lead
to a slight oversubtraction of extended line emission along the line of
sight of the AGN. In our later analysis, we aim at setting lower limits
to the emission line kinematics and fluxes, therefore this procedure is
nonetheless appropriate. In the K-band, we can use the luminous broad
H$\alpha$ emission line as an indicator for how well our algorithm
removes the AGN light. AGN removed spectra are shown in the lower panel
of Fig.~\ref{fig:specnuke} to illustrate the success of the method.

\section{Kinematics and Physical Conditions within the Nebulosity}
\label{ssec:kinematics}

To investigate the spatially resolved kinematics, we extracted spectra
from 3$\times$3 pixel box apertures ($0.75\arcsec\times 0.75\arcsec$;
slightly larger than the estimated seeing in both reduced data
cubes). Using the IRAF task SPLOT, we fitted [OIII]\lam5007 line
emission with up to 3 kinematic components, requiring $S/N > 3 $ in each
component. Systematic errors of these measurements are possibly large,
and the AGN removal in the H band might be not very precise. We thus
use these spectra only to identify zones of rather uniform kinematics,
and extract integrated spectra of these zones to further investigate
their properties.
\begin{figure}[htb]
\centering
\plotone{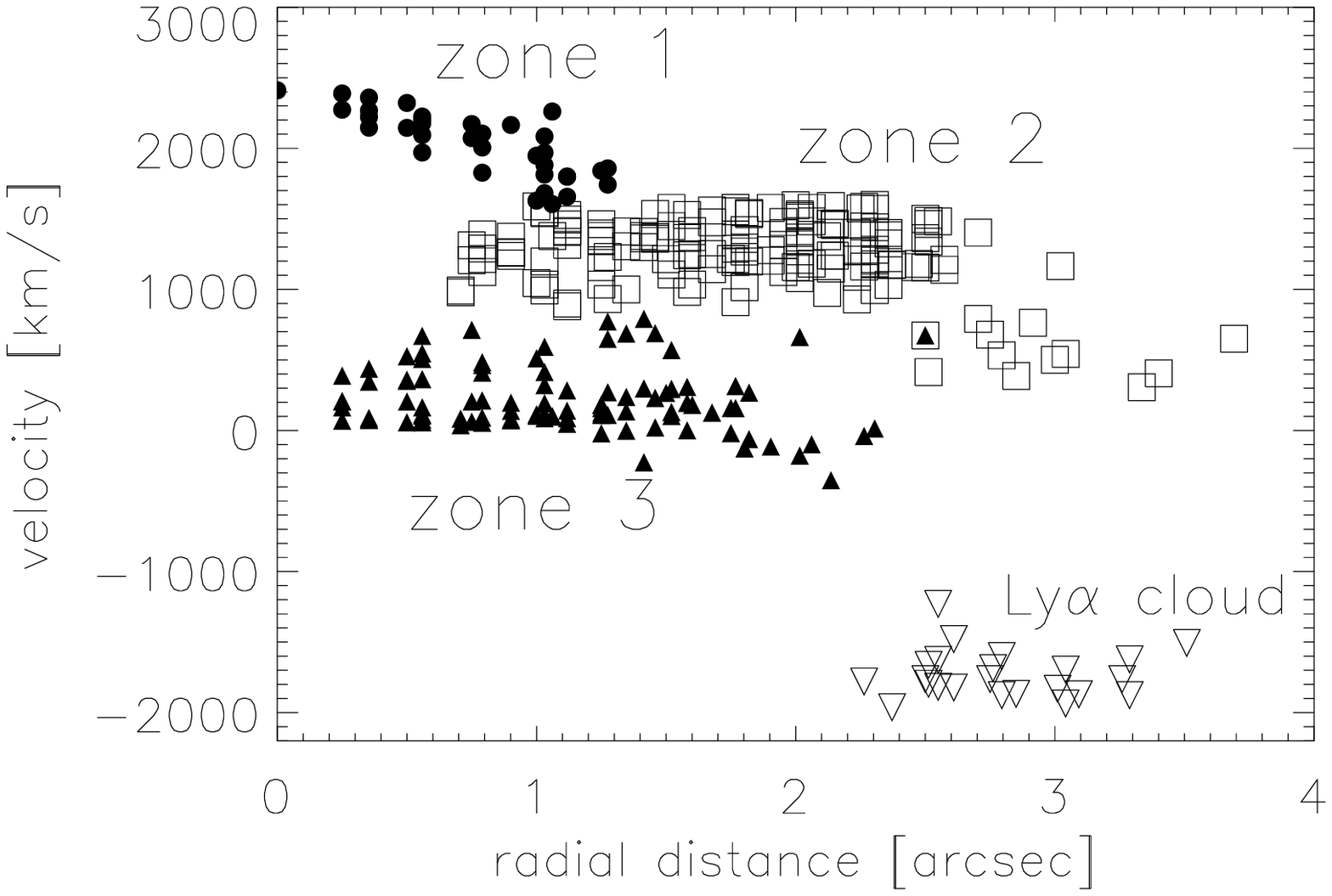}
\caption[Relative velocities vs. projected radius.]{Relative velocities in
the nebulosity of MRC1138-262 as a function of projected distance from the
nucleus. Zero velocity is defined to be the velocity of the highest surface
brightness [OIII]\lam5007 line emission. (Labels 1 to 3 are given to
zones of roughly constant velocity, see Fig.~\ref{fig:velocitymap} for the
spatial distributions of these ``velocity zones''.) ``Ly$\alpha$ cloud''
(upside-down triangles) refers to a zone of extended Ly$\alpha$ emission
defined in \cite{pentericcithesis} and called ``B3''.}
\label{fig:velvsdistplot}
\end{figure}
\subsection{Gas Velocities and Emission Line Widths}

\begin{figure}[htb]
\centering
\plotone{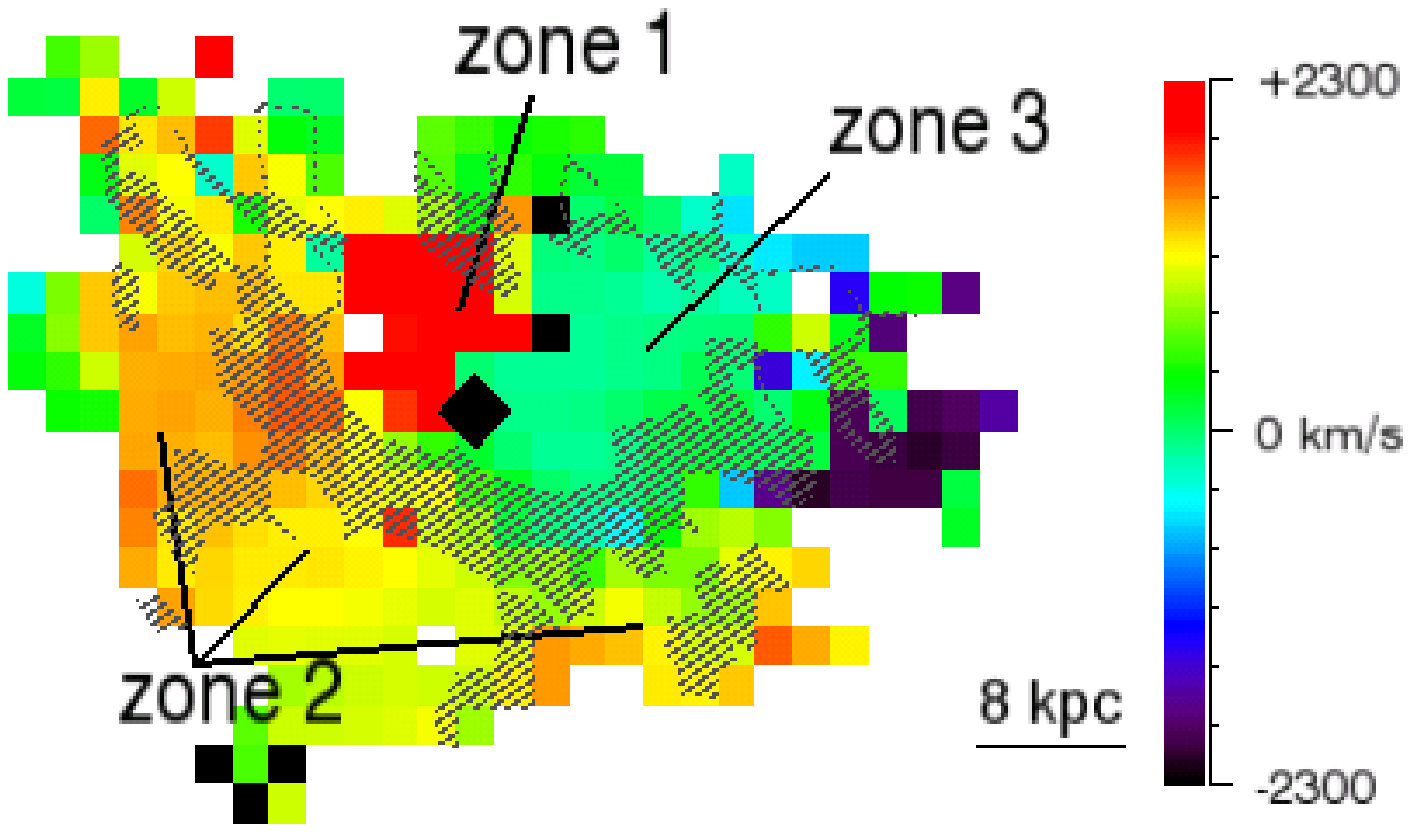}
\caption{The map of relative velocities observed in MRC1138-262 from
fitting the [OIII]\lam5007 emission line. North is up, east to the
left. The zero velocity is defined to be that of the highest surface
brightness [OIII]\lam5007 line emission. The coding is given in the
color bar and the units are \kms.  The hatched area shows the ``ring''
of broad (FWHM$\ge 1400$ \kms) emission lines around zones 1 and
2. The diamond indicates the position of the AGN. }
\label{fig:velocitymap}
\end{figure}

Relative velocities of all [OIII]\lam5007 components are shown in
Fig.~\ref{fig:velvsdistplot} as a function of projected radius. They
appear to be segregated by velocity into roughly 4 groups with
sizes of $R_i = 1-2$\arcsec\ ($8-16$ kpc) over the total radius of
$R_{tot} \sim 2.5-3$\arcsec\ (20-25 kpc).  Overlap between these groups
in velocity is small. In Fig.~\ref{fig:velocitymap} we show the map of
relative velocities of the brightest [OIII]\lam5007 line component in
each spatial pixel.  At least three distinct kinematic zones, each
separated in velocity space by $\sim$ 1000 \kms, but with rather uniform
velocities internally, are evident. Velocities in zone~1 decline with
increasing projected distance from the nucleus, gradually approaching
the values in zone~2 at the eastern bound. This is in agreement with
Fig.~\ref{fig:velvsdistplot}, which also suggests that the velocities
in zone 1 at its low-velocity end approach the velocities in zone
2. Zone 3 has largely uniform intrinsic velocities, whereas velocities
in zone 2 vary gradually with position angle, reaching a  maximum at a
position angle $\sim 230^{\circ}$. The relatively uniform velocities
within each structure, (Fig.~\ref{fig:velocitymap}), mirroring the
regularity found in Fig.~\ref{fig:velvsdistplot}, are highly reminiscent
of what would be expected for overlapping, edge-brightened ``bubbles''
(such a scenario is also favored by the low filling factor of the emission
line gas, \S\ref{sec:timescale} and \S\ref{sec:heatgas}).

However, we are using the term ``bubble'' somewhat loosely, referring 
mainly to the clearly distinct zones in velocity space. Their
morphologies are nearly circular as if projected on the sky, and are
consistent with expanding spheres or projected conical outflows. They
might be analogues of buoyantly rising, low-density regions in
low-redshift radio galaxies, but our analysis does not rely on this
interpretation. By using this term, we simply relate to the geometry and
velocity structure of the gas, not necessarily the full astrophysical
interpretation.

\begin{figure}[htb]
\centering
\plotone{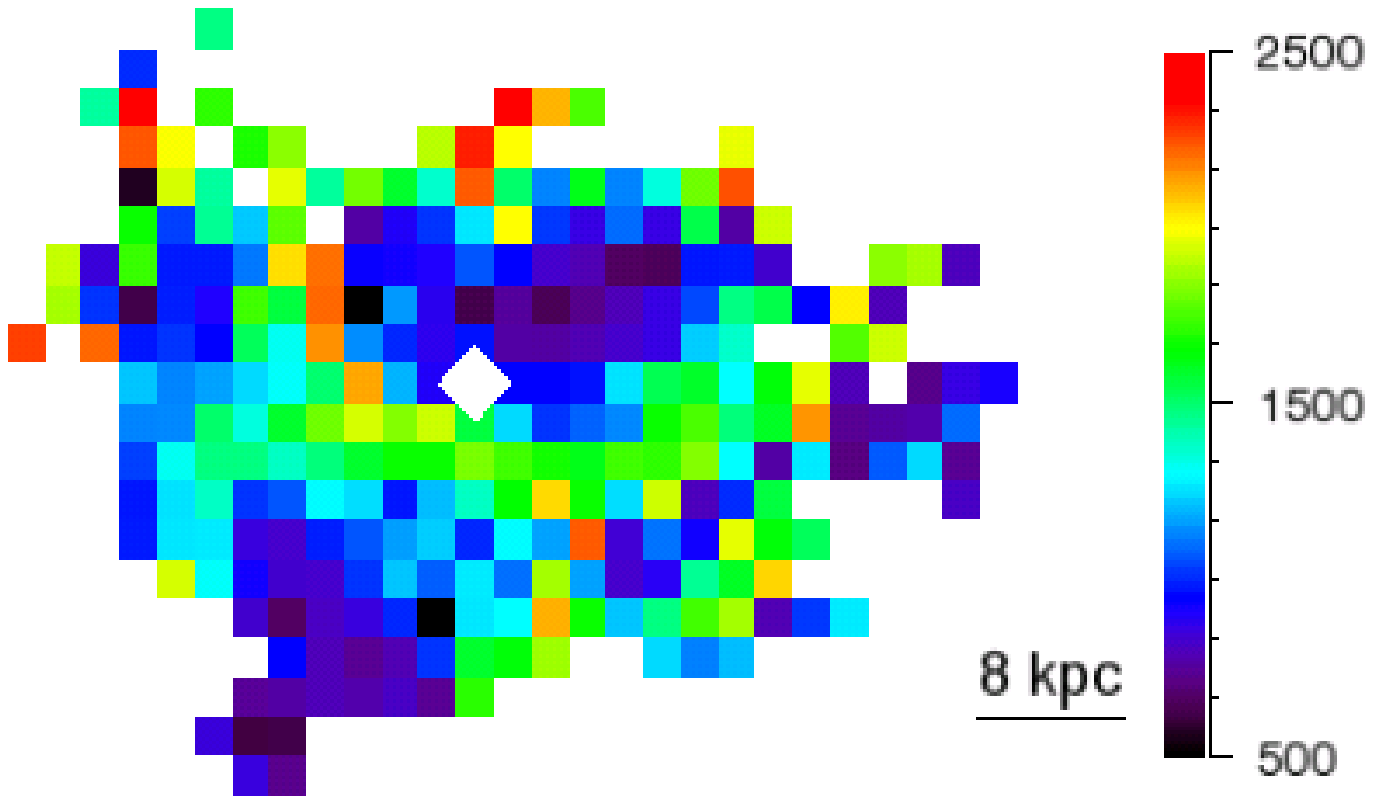}
\caption{The map of full width at half maximum values derived from the
[OIII]\lam5007 emission line.  The color bar on the right shows the values
of FWHM in units of \kms. The diamond indicates the position of the
AGN. North is up, east to the left.}
\label{fig:sigmamap}
\end{figure}

This interpretation is also supported by the spatial distribution of
line widths. In Fig.~\ref{fig:sigmamap} we show a map of the FWHM of
the brightest [OIII]\lam5007 component in each pixel. Line widths vary
between FWHM$= 800-2400$ \kms. Most striking is a ring of broad line
emission (FWHM$\ge 1400$ \kms) around zones 1 and 3, shown as hatched
area in the velocity map of Fig.~\ref{fig:velocitymap}. This ring
supports the above 
interpretation that the emission line regions with well separated
velocities are geometrically and physical distinct entities, which
originate from apparent broadening due to overlap between adjacent
bubbles or from turbulence due to interactions along their
boundaries. 

\subsection{Spectral Properties}
\label{sec:iontempdens}

\begin{figure}[htb]
\centering
\plotone{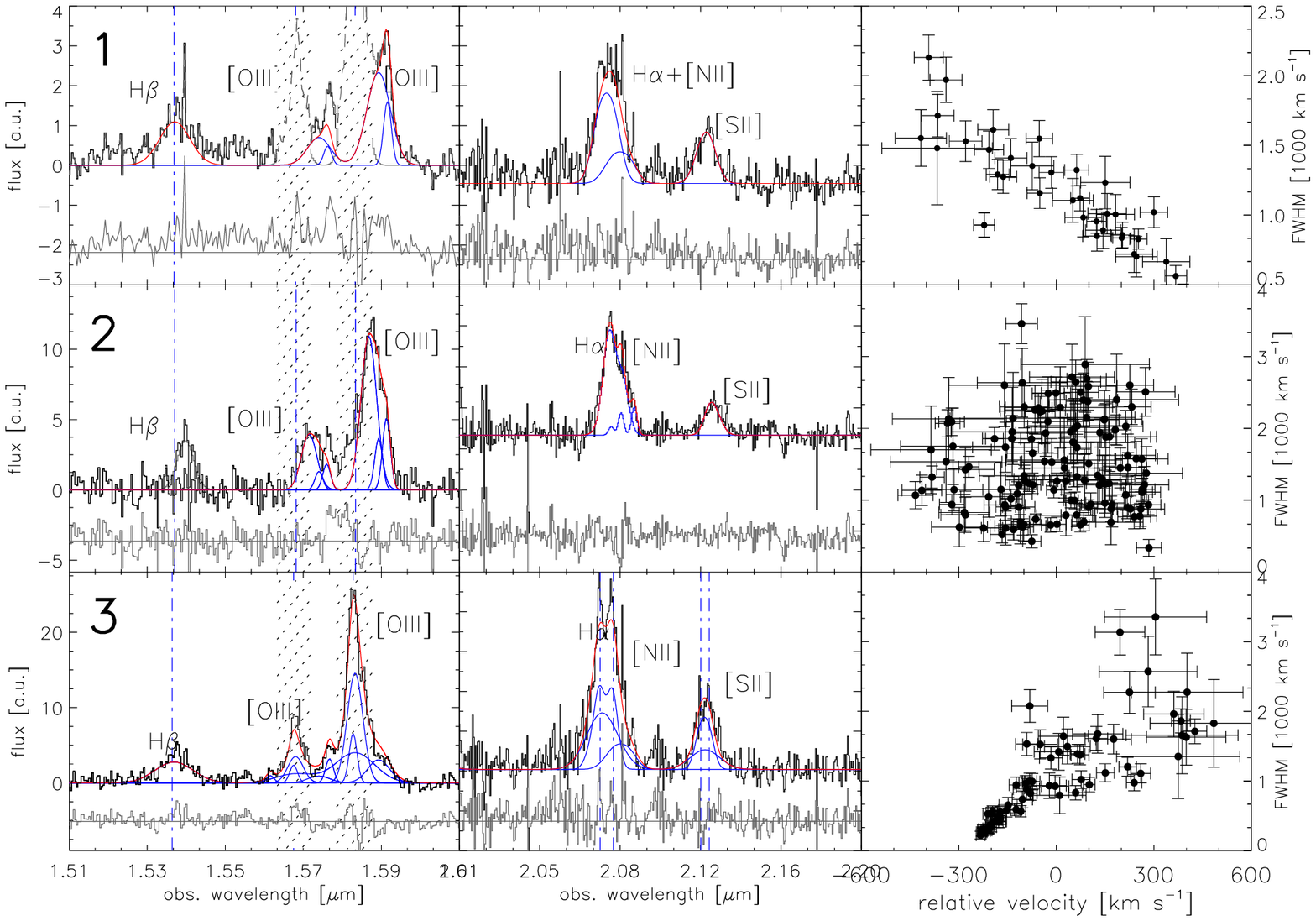}
\caption{{\it left, top to bottom}: Integrated H-band spectra of zones
1, 2, and 3 from Fig.~\ref{fig:velocitymap}, showing the lines H$\beta$,
and [OIII]\lam\lam4959, 5007, respectively.  {\it middle} K band spectra
showing H$\alpha$, [NII]\lam\lam6548, 6583, and [SII]\lam\lam6716,
6731. {\it right} [OIII]\lam5007 FWHM within each bubble as function
of the [OIII]\lam 5007 relative 
velocities within each bubble.  The blue and red lines in each of
the panels on the {\it left} and {\it middle} show Gaussian fits to
individual components and the total line profile, respectively. For
[OIII]\lam5007 the fits represent the components necessary to fit the
full line profile. All other lines were fitted assuming the relative
velocities and widths from the fit to [OIII]\lam5007, leaving the flux in
each component as the only free parameter. We require the ratio of the
[NII]\lam\lam6548,6583 and [OIII]\lam\lam4959,5007 emission line doublets
to be fixed and set by the ratio of their transition probabilities. Fit
residuals (gray spectra) are shifted along the ordinate.}
\label{fig:z13spec}
\end{figure}

Integrated H and K band spectra of the 3 definable bubbles are shown in
Fig.~\ref{fig:z13spec}. Emission lines are bright in all bubbles. Line
profiles are complex, and \ha\ and [NII]\lam6583 are blended in all
zones. To obtain a consistent fit for all lines in a given zone,
we identify kinematic subcomponents from the [OIII]\lam5007 profile,
and restrict all other line fits to have the same relative velocities
and widths. We leave the flux as a free parameter, except for the
ratios of the [NII]\lam\lam6548,6583 and [OIII]\lam\lam4959,5007
emission line doublets, where we fix the line ratios to the ratios of
their transition probabilities.

Our data also include the [SII]\lam\lam6717,6731 line doublet, which
has line ratios sensitive to electron densities in the range of 100
to $10^5$ cm$^{-3}$ \citep{osterbrock89}. We identify two $3\times3$
pixel apertures in bubble 2 where the two components are sufficiently
narrow to be deblended robustly. Line ratios in these apertures are
$R_{[SII]} \equiv F_{\lambda6716}/F_{\lambda6731}= 1.1\pm 0.1$. This
corresponds to a density range $n_e=240-570$ cm$^{-3}$, and $n_e=388$
cm$^{-3}$ at best fit value, assuming a temperature $T=10^4$ K. \citep[The
results of][indicate a most likely temperature range of $1-2\times 10^4$
K in AGN narrow line regions, which is sufficiently accurate for our
purpose.]{kraemer98,heckman79,osterbrock89} \citet{bicknell97} find
densities of $\sim 100$ cm$^{-3}$  in the precursor regions of fast
radiative shocks, constrained by radio observations, in good agreement
with our result using an independent estimate.  The [SII] emission
originates mainly in a ``partially ionized zone'', where the thermal
pressure is $\approx$ few $\times$10$^{-12}$ n$_e$ dynes cm$^{-2}$ and
$n_{tot} \sim 3n_e$ \citep[e.g][]{shull79}.  We use these correction
factors in calculations 
that make use of the total gas density, e.g., the total mass estimate
based on the recombination line flux of Hydrogen. Moreover, we only
use the \ha\ line flux with the AGN contribution removed, as described
in \S\ref{sec:spectra}.

The blending of [NII] and \ha\ complicates the \ha\ flux estimates,
therefore we only aim at estimating robust lower limits to the kinetic
energy in the extended emission line region, and hence minimize the \ha\
contribution in our fits to the [NII]$+$\ha\ line blend, typically
obtaining [NII]$\lam6583/\ha\ =1$. With the adopted cosmology, the
total observed \ha\ flux in all zones corresponds to a total \ha\
luminosity ${\cal L}_{H\alpha} =(14.8\pm 1.2)\times 10^{43}$ ergs
s$^{-1}$. To estimate the ionized gas mass in the diffuse nebula, we
assume case B recombination \citep{osterbrock89}, and our previous gas
density estimates:

\begin{equation}
\nonumber
M_H=\frac{L_{H\alpha}}{h\nu_{H\alpha} \alpha^{eff}_{H\alpha}} m_p n_e^{-1} \\
= 9.73\times10^{8} L_{H\alpha,43} \ n_{e,100}^{-1} \ \msun
\end{equation}

with \ha\ luminosity $L_{H\alpha,43}$ in units of $10^{43}$ ergs s$^{-1}$,
Planck constant $h$, \ha\ frequency $\nu_{H\alpha}$, effective \ha\
recombination coefficient, $\alpha^{eff}_{H\alpha}$, proton mass $m_p$,
and electron number density $n_{e,100}$ in units of 100 s$^{-1}$. With
the electron density of $240-570$ cm$^{-3}$ (\S\ref{sec:iontempdens})
this implies a total ionized gas mass $M_{HII}=2.3-6.5\times10^9$ \msun.

Our analysis can be precise at the order-of-magnitude level
only, and is dominated by astrophysical rather than the
measurement uncertainties. Therefore we will adopt an HII mass of
$M_{HII}=3.7\times10^9$ \msun, corresponding to the measured \ha\
luminosity and electron density at face value and not corrected for
extinction. Thus this estimate is a robust low limit to the intrinsic \ha\
luminosity and total emission line gas mass.

\section{Physical Properties of the Flow}

\subsection{What Drives the Dynamics?}

MRC1138-262 is not the first HzRG known to have broad, spatially resolved
emission line gas \citep[e.g.,][]{mccarthy96,tadhunter91,villar99}, but we
have not found another such galaxy in the literature with integral-field
spectroscopy in the rest-frame optical. For galaxies with spatially
resolved slit spectroscopy, the slit is typically aligned with the radio
axis, so that the morphology of the high-velocity emission line gas and
total energy and mass in the flow are difficult to estimate.

Gas dynamics dominated by the large-scale gravitational potential of
the host galaxy would have smooth velocity gradients and rather uniform
line widths with FWHMs of a few hundred \kms. \citet{villar05}
discuss the relatively quiescent kinematics in the Ly$\alpha$ halo of
the z=2.49 HzRG MRC2104-242, with FWHM$\lesssim 600$ \kms, and argue
that the kinematics might be caused by rotation or infalling gas.  The
kinematics we observe in MRC1138-262, however, are different from
MRC2104-242, quantitatively and qualitatively.

Our integral-field observations are uniquely suited for a more detailed
investigation of what causes the extreme kinematics in MRC1138-262, by
relating the local bulk and turbulent velocities in the gas.  The right
panel of Fig.~\ref{fig:z13spec} shows the FWHMs as a function of velocity
in each bubble. Velocities are relative to the mean velocity in
each bubble. FWHMs and velocities are uncorrelated in zone 2, in zone
1 and 3 they correlate tightly with correlation coefficients $R=-0.84$
and $R=0.87$, respectively. Positive correlations like in zone 3 are
also found in starburst-driven winds, and indicate outflowing gas being
accelerated and becoming more turbulent as the kinetic energy is being
dissipated through shocks \citep{lehnert96a}. Negative correlations like
in zone 1 have been interpreted as deceleration shocks \citep{dopita95},
when gas heated by the AGN collides with the confining, higher density
ISM.

Starburst powered outflows do not appear as good candidates to power
an outflow with the velocity gradients and line widths observed.
Typical outflow velocities in superwinds,
as similarly estimated from optical emission lines, are a few$\times
100$ \kms\ \citep{lehnert96a}.  These velocities are a factor of a few
lower than in MRC1138-262 (not including any correction for projection
effects). The velocities seen in starburst driven winds do not seem to be
a function of either energy injection power or rate \citep{heckman00}
or perhaps redshift \citep{nesvadba06}. \citeauthor{nesvadba06}
find that the outflow observed in the $z=2.6$ strongly star-forming
(SFR$_{FIR}\sim650$ \msun\ yr$^{-1}$) submillimeter galaxy SMMJ14011+0252
does not exceed velocities of $\sim 300$ \kms\ and FWHMs $\sim 250$ \kms.
Most importantly for this study, the line cores in SMMJ14011+0252
are not offset from systemic by large velocities.  In other words,
the overall, large scale kinematics of SMMJ14011+0252 do not appear
to be strongly influenced by the wind, similar to the integrated line
profiles of local starburst galaxies. Moreover, pressures within the
wind are similar to winds in local starburst galaxies, suggesting
similar physical properties, which make it unlikely that the outflow
properties would dramatically change with redshift \citep[see arguments
in, e.g.,][]{heckman00}. Nonetheless, the presence of individual bubbles
and large and irregular velocity offsets in MRC1138-262 do suggest that
some sort of feedback mechanism is driving a massive gas outflow. We
will analyze the properties of this outflow in close analogy to
\citet{heckman90} and show that energy and timescale arguments
strongly favour a wind driven by the AGN as the only plausible
explanation.

\subsection{Timescales, Outflow Rates, and Filling Factor}
\label{sec:timescale}

Our data indicate that the gas reaches large distances, over 25 kpc for
the high surface brightness emission, and might even completely escape
the host potential. A way of estimating the time scale of the flow is to
calculate the crossing time given the observed velocities.  With 
bulk motions of $V = 800-2000$ \kms\ and an average observed
radius of the emission line nebula $R=20$ kpc, the dynamical timescale
is $t_{dyn} = R/V \sim 1-2.5\times 10^7$ yrs, providing a rough
timescale necessary to drive the flow over the observed distances.

If the outflow is powered by the radio source, then this timescale
will be approximately similar to the age of the radio source. With
estimated jet-head advance speeds of $\sim$0.1c \citep[e.g.,][with
ranges of about 0.01c to about 0.2c]{wellman97}, and the ``largest
angular size'' of the radio source at $\nu$=1.41 GHz in MRC1138-262
\citep[11.1\arcsec, or 92 kpc;][]{kapahi98}, we estimate a dynamical
timescale of $\sim$3$\times$10$^{6}$ $f_{proj, jet}$ yrs. Because of
the unknown projection angle, this can only be a lower limit (hence
the factor $f_{proj, jet}$), and the intrinsic timescale is likely a
few times higher, up to a few $\times$10$^{7}$ yrs.  

\citet{wan00} suggest that 
the ages of radio sources decrease with increasing redshift. However,
they use a flux limited sample, so that the average radio power will
increase with redshift, whereas the size decreases, which might at least
in part explain this trend. Using their relationship between radio power
and redshift for the $z=2.2$ of MRC1138-262 implies an age of $\sim$0.4
to 1$\times$10$^{7}$ yrs, similar to our previous estimate.

In either approach, the age of the radio source is not significantly
different from the crossing time of the emission line nebula, and the 
radio jet and lobes reach a factor of $\sim$2 larger radii than the
(detected) high surface brightness emission line gas. This alone
cannot prove that the radio jet is powering the outflow, but it is a
necessary prerequisite. 

To estimate a mass outflow rate, we simply assume that the outflow is
being accelerated with an efficiency that is approximately
constant during 
the jet lifetime. This implies that the current estimate is
representative for the average rate over 
the lifetime of the source. Given the large velocities and size of the
nebula, this assumption is reasonable. This very simple scenario implies
$\dot{M}_{outflow} \sim M_H/t_{dyn} \approx 370$ \msun\ yr$^{-1}$. Of
course, we do expect variations in the instantaneous outflow rate due to
jet precession, variable jet energy or mass density, or perhaps changing
coupling efficiency to the ISM as might be expected for a jet
interacting with a ``clumpy'' surrounding medium. 

Having measured the electron density and relative ionization (assuming
case B recombination), we can also constrain the volume filling
factor of the \ha\ emission line gas. The total observed luminosity
corresponds to a ``total emission volume'' $V_{em}$, which we compare
to the volume of the emission line nebula $V_{n}$ (${\cal O}(10^4$
kpc$^3)$, approximated by the total volume of 3 spherical bubbles
with the observed radii of each zone). This yields a filling factor
$ff_V=V_{em}/V_{n}= 0.15\ L_{H\alpha,43}\ n_{e,100}^{-2}\
V_{n,kpc^3}^{-1}$. We find a filling factor, $\sim 1.6 \times
10^{-6}$, likely signaling that distributed in bright, line-emitting 
sheets, small clouds, and/or filaments. 

\subsection{Energy and Momentum Injection}
\label{sec:injection}
Low filling factors arise naturally when a sufficiently intense,
expanding hot wind sweeps up, accelerates, and ionizes dense hydrogen
clouds within the ISM of the host galaxy. Relative to the lifetime of
the host, an outflow with lifetime $\tau_{AGN}=10^7$ yrs is nearly
explosive, implying impulsive energy injection (but constant over the
lifetime of the event). Following, e.g., \citet{dyson80}, expansion
speed of the shell and energy content of the bubble are related as:

\begin{equation}
v_{shell} \sim 435 \dot{E}_{44}^{1/5} \ n_{0}^{-1/5} \ t_{7}^{-2/5} 
\ {\rm km} \ {\rm s}^{-1},
\end{equation}

where $\dot{E}_{44}$ is the (constant) energy injection rate in units
of 10$^{44}$ ergs s$^{-1}$, $n_0$=0.5 cm$^{-3}$ is the ambient ISM
density in cm$^{-3}$ and $t_{7}$ is the injection time in units of
10 Myrs. We approximate the velocity $v_{shell}$ through the average
of velocities in the line profile fits to the 3 bubbles discussed in
\S\ref{ssec:kinematics}. From the analysis, we find an average projected
velocity of $<v>\sim$ 800 \kms.  The required energy is then
$\sim$4.2$\times$10$^{45}$ ergs s$^{-1}$.

In a second approach, we model the (constant) energy injection rate into
an energy conserving bubble expanding into a uniform medium with ambient
density $n_0$, which serves as a plausible upper bound to the energy
injection rate, 

\begin{equation}
\dot{\rm E}\approx 1.5 \times 10^{46} r_{10}^2 \ v_{1000}^3 \ n_{0.5}
\ {\rm ergs} \ {\rm s}^{-1}, 
\end{equation}

with the radius $r_{10}$ given in units of 10 kpc and velocity $v_{1000}$
given in units of 1000 km s$^{-1}$.  We estimated the half-light radius
of the H$\alpha$ emission, after removing the contribution from the broad
line region, and found $\approx$10 kpc.  For 10 kpc, a flux weighted mean
velocity $v=800$ \kms\ and ambient density $n_{0.5} = 0.5 \ $cm$^{-3}$,
the energy injection rate is $7.7\times 10^{45}$ ergs s$^{-1}$.
This is likely a minimum energy required since the velocities are seen
in projection.  If we use the total extent of the nebula for the radius,
we find the energy injection rate is $\sim 3\times 10^{46}$ ergs s$^{-1}$.

Alternatively, if the bubble is momentum conserving, we estimate that,

\begin{equation}
\dot{p}\approx 4 \times 10^{38} r_{10}^2 v_{1000}^2 n_{0.5} \ {\rm dyn},
\end{equation}

or $2.3 \times 10^{38}$ dyn, of force must be injected to power the
observed flow.  If the jet or radiation field of the AGN is to be a
viable driver of the observed gas dynamics in MRC1138-262, it must
inject at least 10$^{46}$ ergs s$^{-1}$ of energy or alternatively,
about 10$^{38}$ dynes of force into the ambient interstellar medium to
produce the outflow in MRC1138-262.

Each of these estimates is valid within a given model, although the
underlying assumptions are very general, involving basic physical
principles. We can give a fully model-independent, strictly observational
lower bound to the energy injection rate from the observed kinematics
of the \ha\ emission line regions. To this end, we only account for
the observed ionized gas masses, not correcting for projection effects
lowering velocity measurements, extinction, and material in other
phases, especially X-ray emitting or molecular gas, which will contribute
significantly. With the measured \ha\ gas masses and relative velocities,
the kinetic energy is simply $E_{bulk} = 1/2\ \Sigma m_{i} v_{i}^2$
for the bulk velocities and $E_{turb} = 1/2\ \Sigma m_{i} \sigma_{i}^2$
for turbulent motion. For MRC1138-262 and summing over the velocities and
dispersions in the individual bubbles, we estimate $E_{bulk} \sim 2\times
10^{58} f_{proj}$ ergs, and $E_{turb} \sim 3\times 10^{58} f_{proj}$
(with unknown correction for projection, $f_{proj}$). For our dynamical
timescale of $10^7$ yrs, this corresponds to an average injection rate
$\dot{E}_{kin} = E_{turb+bulk}\ \tau^{-1} \approx 1.5\times10^{44}\
t_{10Myrs}^{-1}$ ergs s$^{-1}$ over a lifetime $\tau_{AGN}=10^7$ yrs.

The results illustrate that energy injection rates at high redshift are
not trivial to estimate, which limits the precision of our analysis to
an order-of-magnitude level. It is however encouraging that in spite of
the large uncertainties, the values are broadly consistent. We will in
the following adopt E$_{kin}=1.0\times 10^{46}$ erg s$^{-1}$. This choice
is motivated by the approximate range of values indicated by the simple
estimates we have made and to the order-of-magnitude spirit in which they
were made.

\section{Powering the Outflow: AGN vs. Radio Source}

\subsection{Energy Supply by the AGN}
\label{subsec:mrc.lbol}

If the AGN is to be a viable candidate for powering the observed outflow,
then it must be able to at least provide the above energy injection rate.
It is difficult to know {\it a priori} what is powering the outflow, the
overall bolometric luminosity, ${\cal L}_{bol}$, which might couple to
the ISM through radiation pressure on dust grains, or directly through
the kinetic luminosity of the jet ${\cal L}_{kin,jet}$.  We shall
start by estimating ${\cal L}_{bol}$. Following the scaling of X-ray
to bolometric luminosity of \citet{elvis94} and the X-ray luminosity
of MRC1138-262 at $2-10$ keV, \citep[$\log {\cal L}_{2-10 keV}$=45.6
ergs s$^{-1}$;][]{carilli02}, we estimate $\log {\cal L}_{bol} \sim
46.6\pm 0.6$ ergs s$^{-1}$.  We can also use the relationship between
${\cal L}_{bol}$ and rest-frame optical flux density of the AGN in QSOs
\citep{kaspi00}, and the 5100\AA\ continuum flux in our H band spectrum of
MRC1138-262, $f(5100\AA) =(5.5 \pm 0.53) \times 10^{-17}$ W $\mu m^{-1}$
m$^{-2}$, yielding $\log {\cal L}_{bol}\backsimeq 46.2\pm 0.1$ ergs
s$^{-1}$. Given the approximate nature of either estimate, these values
are consistent. We will in the following adopt $\log {\cal L}_{bol} =
46.6$ ergs s$^{-1}$ since the estimate using the optical continuum is
likely to be strongly affected by high extinction to the region of AGN
continuum emission in radio galaxies compared to QSOs. 

Estimating the jet kinetic luminosity as a measure of the power that
the jet is capable of injecting in the ambient ISM is more difficult
and challenging and many authors have tried a variety of methods for
estimating or constraining the mechanical jet luminosities. We will
therefore use and compare a variety of methods to estimate the
mechanical luminosity of the jet in MRC1138-262. \citet{wan00}
empirically calibrate 
the radio source characteristics in the 3CR as a function of radio power
at 178 MHz. For MRC1138-262 the VLA measurements at 4.86 GHz and the
original selection of the Molonglo Reference Catalog (MRC) at 408 MHz
\citep[][and references therein]{kapahi98}, imply a radio spectral index
$\alpha$=1.34, and a 178 MHz radio power, P$_{178 MHz}$=10$^{34.82}$
ergs s$^{-1}$ Hz$^{-1}$ sr$^{-1}$, in the rest-frame. This corresponds
to a jet kinetic luminosity of 10$^{46\pm 0.3}$ ergs s$^{-1}$ in the
calibration of \cite{wan00}.  \citet{carilli02} measured an integrated
luminosity of the radio jet in MRC1138-262 in the rest-frame 0.1-1.0
GHz band, $10^{45.2}$ ergs s$^{-1}$. Based almost solely on theoretical
arguments, the ratio of mechanical to radio luminosity is typically
argued to be about 10 to 1000 \citep[e.g.,][]{deYoung93, bicknell97}.
Estimating that the radio luminosity is equal to ${\cal O}(10\%)$,
$\epsilon_{10\%}$, of the total jet luminosity sets a lower limit
to the jet kinetic luminosity of 10$^{46.2}$$\epsilon_{10\%}$ ergs
s$^{-1}$. Obviously, it could be substatially higher.  This is, however,
within the scatter of the \citet{wan00} estimate.

We can additionally estimate the jet power using analyses of X-ray
cavities which are thought to be powered by the mechanical energy of
the radio jet interacting with the surrounding hot gas.  Following the
analysis in \cite{birzan04}, we estimate the mechanical luminosity of
the jet in MRC1138-262 to be about 10$^{45\pm3}$ ergs
s$^{-1}$. We think that this estimate is valuable because it is
similar to our approach and provides a robust lower limit to the
mechanical luminosity. However, \citeauthor{birzan04} study radio
sources at lower redshift, which have different radio morphologies
than MRC1138-262 and typically lower radio luminosity. Moreover,
\citeauthor{birzan04} point out that their estimates might be
systematically too low due to various physical effects.  
As a last approach, we can estimate the kinetic energy necessary to
explain the observed expansion speeds of radio lobes which are of the
order of 0.1 to 0.4c \citep[e.g.,][]{conway02}.  These advance
speeds and a plausible model for the medium into which the lobes are
advancing, yields estimated jet kinetic powers of 10$^{46-48}$ ergs
s$^{-1}$ for powerful, high luminosity radio sources
\citep{bicknell03, carvalho02}. 

For the subsequent analysis, we will adopt a finducial jet kinetic
luminosity of 10$^{46}$ ergs s$^{-1}$, but suggest that the true value
could be a factor of a few higher.  Thus, both jet power and
bolometric luminosity estimates provide sufficient energy to power the
observed outflow. Both appear similarly powerful, thus their
relative impact will depend on their coupling 
efficiency with the ambient ISM and their ability to accelerate ambient
interstellar material to large distance from the nucleus.

\subsection{Radiation Driven Winds}
\label{sec:mrcbololumi}

The importance of radiation pressure in driving an outflow depends
on the intensity of the radiation field itself, i.e., the bolometric
luminosity, ${\cal L}_{bol}$. Powerful AGN have radiation fields that
are likely intense enough to efficiently heat and accelerate the gas near
their center through Compton heating \citep{sazonov05} and dust opacity
\citep{king03}. Through Coulomb coupling of charged dust grains to the
ionized gas, radiation pressure can in principle trigger an outflow. Blue
shifted X-ray absorption lines in quasar spectra indeed suggest parsec
scale outflows, with mass loss rates of $\sim$1 M$_{\sun}$ yr$^{-1}$
and velocities $\sim$0.l c \citep[e.g.,][]{reeves03}. However, e.g.,
\citet{king03} find that this gas generally cannot drive large scale
outflows.

We describe a momentum driven AGN wind, following, e.g., \cite{king03} and
\cite{murray05}. The equation of motion for an optically thick wind is,

\begin{equation}
\dot{P}_{rad}=M_g(r)\dot{V}=-\frac{GM(r)M_g(r)}{r^2} + L_{AGN}/c,
\end{equation}

with acceleration $\dot{V}$, and AGN luminosity $L$. $M_g(r)$ denotes
the gas mass as a function of radius. \cite{murray05} argue that the
velocity as a function of the distance for the black hole, $r$, is set by,

\begin{equation}
V(r)=2\sigma \sqrt{\frac{L_{AGN}}{L_M-1} ln(r/R_0)},
\end{equation}

with a threshold luminosity $L_M$ above which the AGN is able to drive
a significant, optically thick outflow, gas fraction $f_g$, velocity
dispersion of the host galaxy $\sigma$, and gravitational constant
$G$. $R_0$ is the radius where the acceleration sets on, namely the
dust sublimation radius at a few pc from the black hole. $L_M$ can be
derived in close analogy to the Eddington limit for the central AGN,
$L_M=4f_g\ c \sigma^4/G$. \citep[See \S 5 of ][ for the full derivation
of this threshold.]{murray05}. For a $\approx {\cal L}^{\star}$
galaxy, $L_M\backsimeq 3\times 10^{46} \ f_{g0.1}\ \sigma_{200}^4$
ergs s$^{-1}$.  With ${\cal L}_{bol}=1-4\times 10^{46}$ ergs s$^{-1}$,
a reasonable gas fraction of 10-50\%, and a velocity dispersion of 250
km s$^{-1}$ (roughly corresponding to the stellar mass of MRC1138-262),
this suggests that $L_M\backsimeq 10^{47}$ ergs $s^{-1}$, well above
the bolometric luminosity of MRC1138-262. For a lower gas fraction and
smaller velocity dispersion, radiation pressure might play a role, but
it unlikely dominates in accelerating the gas to the high velocities we
observe. For instance, if $L/L_M$=$\frac{4}{3}$ \citep[in good agreement
with the $L/L_M\approx 1$ implied by ][]{murray05}, the acceleration
would have to have been maintained for a few $\times$ 10$^{8-9}$ yrs
in order to accelerate the emission line gas in MRC1138-262 to the
observed velocities, likely about 1-2 orders of magnitude longer than
the AGN lifetime. Typical line velocities would be about $1/2$ those
observed, perhaps less since we cannot accurately quantify the effect
of projection. Hence, radiation pressure does not seem sufficient to
power the large-scale gas kinematics in MRC1138-262, making the radio
jet a more likely candidate to mediate the AGN feedback.

\subsection{Coupling Efficiency between Jet and ISM and the ``Dentist
Drill'' Model}

We derive a simple estimate for the observed coupling efficiency $\eta=
E_{kin}/E_{in}$ between the energy input rate $\dot{E}_{in}\sim 1\times
10^{46}$ ergs s$^{-1}$ of the radio jet (\S\ref{subsec:mrc.lbol}) and the
energy injection rates derived in \S\ref{sec:injection}, $\dot{E}_{kin}=
(0.5-1) \times 10^{46}$ erg s$^{-1}$. Given the uncertainties in either
estimate, and in the spirit of providing lower limits, we adopt a loosely
constrained coupling efficiency $\eta = {\cal O}(10\%)$, corresponding
to the jet kinetic energy estimate common to all methods of
\S\ref{subsec:mrc.lbol}.

The ``dentist drill model'' \citep{scheuer82} is the simplest scenario
explaining how the jet might interact with the ambient ISM, which uses a
termination shock to couple with the ISM as the jet-head ploughs through
the ambient medium. Small variations of the jet direction cause the jet
to jitter across the working surface, and the resulting cavity will be
somewhat larger than the jet diameter itself. Typical assumed values
are $\sim 15^{\circ}$ \citep{begelman89}.

Thus in the ``dentist drill'' scenario, the interaction between radio jet
and the surrounding medium is limited to the working surface of the jet,
and the coupling efficiency is approximately the area of the working
surface, divided by 4$\pi$. This can be parameterized as $\epsilon (\%)
= (A_{jet}/4\pi) = 0.003\% A_{sq degree}$, or 0.045\% for 15\mydeg\
opening angle. To explain the $\sim 10^{46}$ erg s$^{-1}$ kinetic
energy injected into the outflow of MRC1138-262 with the small
$\epsilon$ implied by the dentist drill model would hence require an 
energy input of $>$ few $\times 10^{49}$ erg s$^{-1}$. This is above
even the most extreme estimates of jet kinetic energies discussed in
\S\ref{subsec:mrc.lbol}. Hence the observed kinematics in MRC1138-262
require a significantly more efficient coupling to the ambient ISM
than implied by the ``dentist drill'' model.

\section{Heating the Gas: Shocks vs. Direct Illumination by the AGN}
\label{sec:heatgas}

Line widths and bulk motions in MRC1138-262 indicate shock speeds $\gtrsim
500$ \kms. Optical emission line properties of shocks with such speeds
are very difficult to model, because emission arises predominantly
through ionizing photons escaping upstream ahead of the shock into the
``precursor'' region.  Thus, accurately modelling the line emission from
the shock depends crucially on the detailed properties of the ambient ISM
and the radiation field (both background and that generated by the shock).
Following this argument and the modeling of \citet{dopita96}, we use,

\begin{equation}
\dot{M}_{shock} \approx 4.4 \times 10^4 \ n_{2\gamma\ per\ H}^{-1}\
L_{H\alpha,43}\ \msun
\end{equation}

to estimate the amount of shocked material from the measured H$\alpha$
luminosity in MRC1138-262 (\S\ref{sec:iontempdens}). In this equation,
$n_{2\gamma\ per\ H}$ is the number of recombinations per Hydrogen
atom. At least $2.5 \times 10^5$ \msun\ yr$^{-1}$ of shock material are
necessary to produce the entire H$\alpha$\ luminosity, not accounting
for extinction and possibly a lower efficiency of the emission (fewer
recombinations per Hydrogen atom for example).  Over the likely $10^7$
yrs lifetime of the radio jet (\S\ref{sec:timescale}), this implies an
unphysically large $>10^{12} M_{\sun}$ of shocked gas.

On the other hand, the predicted surface brightness of a shock with
500 km s$^{-1}$ velocity is about 2$\times$10$^{40}$ ergs s$^{-1}$
kpc$^{-2}$ \citep{dopita96}, and increases dramatically with velocity
($\propto V_s^{2.4}$). Shock heating across the projected area of the
emission line nebula in MRC1138-262, $\sim 1200$ kpc$^2$, would therefore
produce a total H$\alpha$\ luminosity of $\sim$2.4$\times$10$^{43}$,
$\lesssim 15$\% of what we observe. Accounting for extinction will make
this discrepancy yet more dramatic.

Since the amount of shocked material necessary to power the nebula is
unphysically large and the predicted surface brightness is too low, it
seems that shock excitation is probably not viable as the sole excitation
mechanism of the emission line gas, and ionization by the AGN will play
a large role. \citet{villar03} reach a similar conclusion from
longslit rest-frame UV spectra of 10 HzRGs.

To estimate whether photoionization by the AGN can heat the gas, we assume
a uniformly dense, radiation bounded nebula in ionization equilibrium.
With the properties we observe in MRC1138-262, this requires an ionizing
photon rate of $Q(H^0) \approx 3.7\times10^{54}\ L_{H\alpha, 43}
s^{-1}$, or $5.5\times10^{55} s^{-1}$ for $L_{H\alpha, 43}=14.8$. 
The results of \cite{elvis94} imply that on average each ionizing
photon in powerful AGN has an energy corresponding to $\sim 3$ times
the ionizing potential of hydrogen (i.e., $E_{\gamma}=3\times 
h\ <\nu>_{ion}$). We therefore constrain the ionizing luminosity $L_{ion} =
3\ Q(H^0) h<\nu>_{ion}$, corresponding to $2.4\times10^{44}\ L_{H\alpha, 43}\
ergs\ s^{-1}$, or $\approx 3.5\times10^{45}\ ergs\ s^{-1}$. If $\sim 10$\%
of the bolometric luminosity of MRC1138-262 is ionizing \citep{elvis94},
i.e., $\sim 4-6\times 10^{45}$ ergs s$^{-1}$ (\S\ref{subsec:mrc.lbol}),
the nebulosity of MRC1138-262 might well be mainly photoionized by the
AGN, if the escape fraction is  $\lesssim 90$\%. This estimate also
implies that the nebula could easily be matter bounded.

However, given the small filling factor ($\sim 10^{-6}$,
\S\ref{sec:timescale}), a covering factor $\sim 10$\% implies that
the gas is distributed in elongated structures with large surface
area and relatively small volume. The gas must either be clumped into
large numbers of small, approximately spherical clouds, long and thin
``string-like'' structures, or in thin, broad sheets, for either shock
or excitation of the nebula.  Whatever the exact geometry, our results
are only consistent with a small volume filling factor, but a relatively
large covering fraction.

Thus, the mechanical and optical emission line properties of MRC1138-262
give direct evidence for an outflow which is driven by the expanding radio
source, and heated mainly through photoionization by the AGN.

\section{AGN Feedback: Impact on Galaxy Evolution and Cosmological
Significance}\label{sec:significance}

Our analysis of the kpc scale outflow in MRC1138-262 suggests
that the AGN is the best candidate for driving the outflow: The
crossing time of the nebula ($\sim 10^7$ yrs) approximately equals
the lifetime of the radio jet and typical estimates of AGN lifetimes
\citep[][and references therein]{martini04}, and the AGN also provides
the energy input rate necessary to power the observed outflow, given
it is efficiently coupled to the ISM, with ${\cal \epsilon}= {\cal
O}(0.1)$. Observed mass loss rates are $\gtrsim 400$ \msun\ yr$^{-1}$
(\S\ref{sec:timescale}). Population synthesis fits assuming exponentially
declining star-formation histories with a variety of e-folding times
(including infinity for continuous star-formation) over a wide range
of maximum star-formation rates, suggest a stellar mass of $\sim
5\times 10^{11}$ \msun. The range of best fit masses for all of the
models suggest an uncertainty of about a factor 2-3. The observed
outflow rate therefore corresponds to a mass loss of $\sim 10$\% of
the stellar mass within $10^8$ years. Observations of carbon monoxide
in high redshift radio galaxies suggest gas masses of $\sim 10^{10-11}$
\msun\ \citep{papadopoulos00,greve05}, corresponding to $\lesssim 20$\%
of the stellar mass of MRC1138-262. Given the large uncertainties in
these estimates, we assume a gas fraction of $\lesssim 50$\% of the
stellar mass, implying
that AGN feedback could remove the total ISM of the host within $\sim
5\times 10^8$ yrs. We emphasize that the mass loss rate is a strict lower
bound estimated from the observed \ha\ emission line gas mass. The true
outflow might be up to an order of magnitude higher, and might well be
larger than the total needed to quench future star-formation.

The outflow in MRC1138-262 appears not aligned with the radio jets,
but has a nearly spherical geometry, so that an adequate assessment
of the dynamics in the extended emission line region clearly requires
integral-field data, especially in the rest-frame optical, to be less
sensitive to the source geometry, extinction, and the complex radiation
transport in the rest-frame UV given the complex spatial structure of the
emission line gas. We do not believe that MRC1138-262 is a ``one-off''
special case. Broad forbidden emission lines in the nebulosities of
HzRGs with FWHM$ > 1000$ \kms\ are commonly found in longslit studies
\citep[e.g.][]{tadhunter91,mccarthy96,iwamuro03}. Simulating
a longslit spectrum of MRC1138-262 from our SPIFFI data with the
``slit''aligned with the radio axis, yields similar \ha\ fluxes as given
in, e.g. \citet{evans98}. We recently obtained SINFONI data of a small
sample of HzRGs at z=2.5-3.5, and find similarly extreme kinematics in
the extended sources. In the following we will therefore assume that
MRC1138-262 is in an evolutionary stage, which is short, ${\cal O}(10^7)$
yrs, but common for massive galaxies, and in particular for those which
go through a phase of powerful nuclear activity at high redshift.

\subsection{A Solution of the [$\alpha$/Fe] Puzzle?} 
\label{subsec:alphafe}

A massive, AGN driven outflow is a catastrophic event which can be
expected to leave measurable imprints on the low-redshift descendents of
the AGN host galaxy. Semi-analytical models of massive gas-rich mergers
with a multi-phase ISM in the early universe are found to reproduce
the local luminosity function of massive, red early-type galaxies if
they include a phase of strong AGN feedback \citep[][]{hopkins05,
croton06}. \citet{robertson05} find that these models can explain
the tilt in the fundamental plane assuming dissipative, high redshift
merging of galaxies with gas fractions $>0.3$, including star-formation
and AGN-related feedback, and \citet{dimatteo05} are able to reproduce
the observed M$-$$\sigma$ relationship using this mechanism. In
spite of these successes, such models cannot resolve the
black hole accretion, therefore AGN feedback is included in a rather
schematic way, neglecting,e.g., radiatively inefficient black hole growth
\citep{springel05}. They assume isotropic thermal coupling between AGN and
ISM with efficiencies of $\sim 0.5$\%, and imply that star-formation is
suppressed within a few $\times 10^8$ yrs. We note that this efficiency
is lower than the coupling efficiency we find in MRC1138-262 (${\cal
L}_{AGN}/E_{flow} \gtrsim {\cal O}(0.1)$), but the time scales agree
surprisingly well, given the crudeness of the models and, by necessity,
the ``order-of-magnitude'' nature of our analysis.

To set observational constraints on the impact of AGN driven winds, we now
investigate whether the outflow in MRC1138-262 can explain the observed
properties of the low redshift massive galaxy population. A massive
outflow related to major mergers at high redshift is in 
overall agreement with a population of mostly spheroidal galaxies,
stellar populations that have luminosity-weighted ages consistent with
formation at redshifts $z\sim 3-5$ \citep{thomas05}, and with extremely
low gas contents.

A more quantitative constraint comes from the relative enhancement of
$\alpha$ elements with respect to the iron content of these galaxies,
$[\alpha/Fe]$ \citep[e.g.,][]{thomas99,romano02}. The abundance of
$\alpha$-elements relative to iron traces the relative numbers of
thermonuclear and core-collapse supernovae and effectively traces the time
scale of star formation due to the relative delay in the thermonuclear
supernova. \citet{silk05} propose a top-heavy IMF as alternative
explanation for the observed $[\alpha/Fe]$ overabundance. However, as
\citet{trager00} point out, the {\it relative} $[\alpha/Fe]$ overabundance
in fact is an {\it absolute} Fe underabundance, which indicates a relative
lack of contribution of the thermonuclear supernovae to the overall metal
abundances rather than an enhancement of core-collapse supernovae.  This
argument does not favor explaining the relative $[\alpha/Fe]$
enhancement compared to the solar value with a top-heavy IMF. Moreover,
\citet{nagashima05} find from semi-analytical modeling that the hypothesis
of a top-heavy IMF cannot explain the observed tight correlation between
$[\alpha/Fe]$ enhancement and velocity dispersion in early-type galaxies,
but agrees with a more pronounced early truncation of star-formation
with increasing galaxy mass.

A mechanism which efficiently removes nearly all gas from the host in
$\lesssim 0.5-1$ Gyrs is therefore more promising, and was suggested
by \citet{thomas99} and \citet{romano02} for massive early-type
galaxies. Outflows with the observed mass loss rate in MRC1138-262,
$\dot{M} \gtrsim 400$ \msun\ yr$^{-1}$, fulfill this constraint, given
that sufficient kinetic energy is injected into the ISM during the AGN
luminous phase.  In \S\ref{subsec:simpleenergy} we estimate that $\sim few
\times 10^{60}$ ergs or more are injected into the ISM of MRC1138-262
during an assumed AGN lifetime of $10^7$ yrs. The mechanical energy of
the AGN is able to entrain ambient material, propelling it outwards at
the high velocities observed. The maximum amount of ambient material
(if accelerated to escape velocity) can be approximated by an energy
conservation argument as:

\begin{equation}
M_{ej,max} = 2 \times 10^{11} E_{outflow,60} \ v_{esc, 500}^{-2}\ \msun
\label{eqn:Mejmax}
\end{equation}

where $E_{outflow,60}$ is the energy of the outflow in units of
$10^{60}$ ergs and $v_{esc, 500}$ is the escape velocity in units of 500
\kms. $v_{esc}$ is difficult to constrain. For galaxies with NFW profiles
\citep{navarro97}, there is a simple relationship between $v_{esc}$,
virial radius and mean mass density of the halo over the background dark
matter density \citep[see][]{bullock01}. For a baryonic mass fraction of
10\%, the escape velocity in MRC1138-262 is about 700 \kms\ at the virial
radius of about 120 kpc \citep{bullock01}.  Equation~\ref{eqn:Mejmax}
implies that MRC1138-262 may eject, $M_{ej,max}^{MRC1138-262} \approx$
9 $\times$ 10$^{11}$ \msun\ of material (assuming $E_{outflow,60}$=9.,
which is the maximum of our estimated values, and $v_{esc, 500}$=1.4),
or $M_{ej,max}^{MRC1138-262} \approx$ 3 $\times$ 10$^{11}$ \msun\
assuming our fiducial value of $E_{outflow,60}$=3..  This is of course
a very crude estimate, but it does suggest that, plausibly, much of the
baryonic mass could escape the potential of the galaxy and thus will
not be available for future star-formation.  However, this represents
a strict upper limit to the total mass ejection.  We have not assumed
a coupling factor between the energy and the ambient gas, and this
factor is undoubtably less than one (we estimate ${\cal O}(10\%)$).
Subsequently, we will adopt this maximum mass ejection estimate and
a coupling efficiency of 10\% as the upper boundary, and the simple
observed mass loss rate from \ha\ as a lower boundary to the most likely
range of mass loss rates and to constrain the amount of mass and metals
ejected by AGN winds.

\subsection{Can AGN Inhibit Baryonic Mass Accretion?} 
\label{subsec:virialinfall}

In addition to suppressing star formation within the short timescale set
by the [$\alpha$/Fe] ``clock'', the AGN must also be able to inhibit the
accretion of gas from the surrounding IGM or ICM to effectively halt
future star formation. \citet{croton06} find that it is necessary to
incorporate such a mechanism into their simulation, to inhibit galaxy
growth in a population of massive red galaxies, and to produce a mass
function with the appropriate exponential cutoff at the high end. In
a very interesting analysis of a large sample of radio loud AGN based
on data from the Sloan Digital Sky survey, \citet{best05} find that the
dependence of the size of the radio-loud AGN fraction on black hole mass
mirrors that of the cooling rate within the hot atmospheres of elliptical
galaxies. \citeauthor{best05} further argue that the AGN could be fueled
by such cooling gas and that AGN heating could plausibly balance the
cooling of the gas over time. Could the energy output of MRC1138-262
balance gas accretion from the IGM with a similar mechanism?

Following the arguments in \citeauthor{croton06}, we can estimate whether
or not it is even plausible for the radio jets in MRC1138-262 to inhibit
the accretion of gas onto the dark matter halo.  \citeauthor{croton06}
estimate the rate at which gas is accreted onto dark matter halos as
a function of halo virial mass and redshift.  For a halo with $V_{vir}
\sim 250$ \kms\ at $z=3$, they give an accretion rate of $\sim 200-300$
\msun\ yr$^{-1}$.  The gas is assumed to form into a hot, cooling halo
of gas filling the dark matter halo, thus its energy can be estimated
using simple kinetic arguments. Making this assumption, the rate at
which energy is deposited in the halo with a virial velocity, $V_{virial}$,
is approximately:

\begin{eqnarray}
\nonumber
\dot{E}_{cool, virial} \approx \frac{1}{2} \dot{m}_{cool} V_{virial}^2
= \\ 6 \times 10^{42} \left({\dot{m}_{cool}\over{100\ \msun {\rm
yr}^{-1}}}\right) \ \left({{V_{virial}}\over{250\ \kms}}\right)^2 {\rm
ergs}\ {\rm s}^{-1}
\end{eqnarray}

Over the lifetime of the AGN, this will be, $E_{total,cool}\approx
\dot{E}_{cool,virial} \tau_{AGN} \lesssim 6 \times 10^{57} {\rm ergs}$.
Thus both the instantaneous ($\ga$10$^{46}$ ergs s$^{-1}$) and the
total energy ejection rate ($\ga$3 $\times$ 10$^{60}$ ergs) of the AGN
appear sufficient to stop the collapse of gas at the virial radius of
the dark matter halo.  If there is a connection between mass accretion
and igniting AGN activity in galaxies, then the output of a single radio
loud AGN cycle appears sufficiently energetic to suppress subsequent gas
flow at the virial radius. Radio lobes at high redshift reach distances of
$\gtrsim 100$ kpc. Therefore the range and amount of energy injection seem
sufficient to plausibly resist the gas accretion flow, and to influence
the final baryonic mass of galaxies hosting powerful radio sources.
Of course the energy output from the AGN must couple to the infalling
gas for it to be effective.  Certainly the high efficiencies we have
estimated for the gas within 10-20 kpc of the nucleus would be sufficient.

This is a fundamental difference to starburst-driven winds which will heat
the gas on similar time scales, but lack the energy and power necessary
to unbind significant amounts of gas from the most massive halos, and
are very unlikely to stop the accretion flow. If the ejected gas rains
back onto the galaxy, and this material supports further star-formation,
the enhanced $[\alpha/Fe]$ enhancement will not be preserved. Similar
arguments disfavor models where star-formation has consumed most of
the gas prior to the AGN bright phase, e.g., \citet{sazonov05}, and
approaches where ultraluminous starbursts are triggered through {\it
positive} AGN feedback \citep{silk05}, provided that most of the stellar
mass observed at $z=0$ are formed in such a burst.

\subsection{The Cosmological Significance of AGN-feedback}
\label{subsec:simpleenergy}

\subsubsection{Simple Energetic Argument}
\label{subsubsec:simpleenergy}
The short lifetimes of powerful AGN \citep[$\sim 10^7$ yrs;][]{martini04}
compared to a Hubble time indicate that nuclear activity is likely
episodic. To provide an estimate of the cosmic energy density released
through AGN feedback, we will therefore assume that AGN hosts have only
one cycle of activity that lasts $\tau_{AGN} \sim 10^7$ yrs with coupling
efficiencies to the host's ISM and outflow characteristics similar to
MRC1138-262.  The energy output during one such cycle follows from the
previous analysis, $E_{tot,1138}\approx\dot{E}_{1138}\times\tau_{AGN}\sim
3 \times 10^{60}$ ergs (with $\dot{E}_{1138}$=10$^{46}$ ergs s$^{-1}$ and
$\tau_{AGN}$=10$^7$ years).  To estimate the AGN duty cycle directly from
the number of observed luminous AGN, we use the co-moving number density
of luminous AGN calculated by, e.g., \citet{pei95}, $\Phi(QSO) \approx
5\times 10^{-7}$ Mpc$^{-3}$ for optically selected high-redshift quasars
brighter than $M_B=-26$. If the ``unified model'' of AGN is correct,
then QSOs are a subsample of all AGN, and the observed number density
of QSOs will be a few times lower than the intrinsic number density of
the full AGN population. We also assume that a ``radio loud'' episode
is an intrisic part of the lifecycle of all powerful AGN.  Figure~9 of
\citet{pei95} shows that 95\% of all quasars are at $z_{QSO}=1.4-4.0$,
within an epoch of $\sim 2.9$ Gyrs. For $\tau_{AGN}=10^7$ yrs, this
yields $f_{duty} \sim 300$. This is again a lower limit, and neglects
radiatively inefficient black hole growth.

The overall energy budget of powerful AGN can then be gauged by, 

\begin{equation}
\label{eqn:agndens1}
\frac{dE_{AGN}}{dV} = \Phi_{QSO} \ f_{duty} \ E_{tot,1138}.
\end{equation}

As a result, powerful AGN provide a cosmic energy density due to outflows
of $\gtrsim 4.5\times10^{56}\ {\rm ergs} \ {\rm Mpc}^{-3}$. We note that
this {\it global} estimate is relatively robust against uncertain AGN
lifetimes, because increased durations leads to lower duty-cycles and
vice versa.

\subsubsection{Local Black Hole Mass Density}

We can also use the local co-moving mass density of supermassive black
holes (SMBH) as a measure for the integral accreted black hole mass
to constrain the cosmological significance of AGN feedback. Using the
co-moving number density of spheroids and the M$_{BH}-\sigma$ relation,
\cite{yu02} estimate the local mass density of nuclear black holes,
$\rho_{SMBH}\sim 2.9\pm 0.5 \times$ $10^5$ h$_{70}^2\ {\rm M_{\rm
    \sun}}$ {\rm Mpc}$^{-3}$.
To estimate the total kinetic luminosity corresponding to this black
hole mass density, we use the simple scaling,

\begin{eqnarray}
\label{eqn:agnkinlum}
\nonumber
\rho_{E,kin}=\epsilon_k \epsilon_{rad} \Delta M_{BH} c^2 = \\
5.2 \times 10^{57} \epsilon_{k,0.1} \epsilon_{rad, 0.1}\ {\rm ergs}\ 
{\rm Mpc}^{-3},
\end{eqnarray}

with the efficiency of the kinetic energy output relative to the radiative
(bolometric) luminosity, $\epsilon_{k,0.1}$, and the radiative efficiency
relative to the accreted rest mass energy, $\epsilon_{rad}$. The radiative
efficiency of massive black holes has been estimated to be roughly 10\%
to 30\% \citep[e.g.,][and references therein]{yu02}, whereas there
is a broad range of relative kinetic energy efficiency of 0.05 to
1 \citep{willott99}.  Adopting this range of values suggests that the
kinetic energy density is approximately 2 to 100 $\times$10$^{57}$ ergs
Mpc$^{-3}$. For a fiducial $\epsilon_{k,0.1}$=1, the estimated kinetic
energy density ejected is about 5 $\times$ 10$^{57}$ ergs Mpc$^{-3}$,
$\sim 10\times$ higher than what we estimated from the QSO co-moving
space density and duty cycle.

This however implies that all black hole growth is related to gas
accretion, and occurs during AGN luminous phases. Both assumptions are
unrealistic, as indicated, e.g., by the large population of obscured
quasars and the possibility of BH-BH mergers. This result therefore
sets an upper constraint to the energy density ejected by AGN. In the
spirit of setting lower constraints we adopt the first value, and note
that our calculation does not violate any low-redshift constraints from
the low-redshift number densities of supermassive black holes.

\subsection{The Influence on the IGM: Overall Energy, Mass, and Metal
Ejection} \label{subsec:enermassej}

Simple energetic arguments alone show the potentially large significance
of AGN feedback on the evolution of $\gtrsim {\cal L}^{\star}$ local
early type galaxies.  \citet{bernardi03} estimate a local co-moving space
density of ${\cal L}^{\star}$ galaxies, 0.0020 h$^3_{70}$ Mpc$^{-3}$,
\citet{sheth03} find a local co-moving space density of 0.0022 h$^3_{70}
$ Mpc$^{-3}$ for M$^{\star}$ early type galaxies.  Using the co-moving
density of M$^{\star}$ early type galaxies as a normalization factor,
and the upper limit to the kinetic energy density in
\S\ref{subsubsec:simpleenergy}, $9\times 10^{60}$ ergs Mpc$^{-3}$,
we find that {\it on average}, each M$^{\star}$ early-type galaxy ejects
$\langle E_{ej}\rangle \lesssim 6\times 10^{59}$ ergs, roughly the
binding energy of an elliptical galaxy with total mass $M_{tot}\sim 4
\times 10^{11}$ \msun \citep[or a galaxy about one magnitude brighter than
${\cal L}_r^{\star}$;][]{bernardi05}. This is of course a statistical
estimate to relate the impact of such outflows to the population of
massive galaxies. It does not require or imply that each M$^{\star}$
early-type galaxy indeed went through a MRC1138-262-like epoch.

Given the large amount of ambient ISM that the AGN likely drives out
from the galaxy and perhaps from the surrounding dark matter halo,
it is pertinent to ask: If MRC1138-262 is typical, how much mass (and
mass of metals) could such outflows eject?  Using either the observed
outflow rate for $10^7$ yrs or the maximum total mass ejection, we have,
$\Delta M = 0.4-9 \times 10^{10}$ \msun\ (\S\ref{subsec:alphafe}), with
the co-moving QSO number density  and the inverse of their duty cycle,
we find that on average,

\begin{equation}
\rho_{ej} \gtrsim 0.6-14 \times 10^{6} \Phi_{QSO, -6.3} \ f_{300}\ \msun\  {\rm Mpc}^{-3} 
\end{equation}

of material will be swept out by AGN feedback, where $\Phi_{QSO, -6.3}$
is the co-moving space density of QSOs in units of 5$\times$10$^{-7}$
Mpc$^{-3}$ and $f_{300}$ is the inverse duty cycle in units of 300. This
implies that approximately $3-70 \times 10^{8}$ \msun\ of material is
ejected per M$^{\star}$ galaxy. Within these loose constraints, this is
about the mass of ISM in a low-redshift gas rich galaxy and also roughly
the molecular gas mass in high-redshift submm and radio galaxies.

Lacking robust constraints on the gas-phase metallicity in HzRGs, we
use the metallicities of low-redshift massive galaxies to constrain
the metal loss induced by AGN feedback. The local mass-metallicity
relationship indicates that local $>{\cal L}^{\star}$ galaxies have solar
and greater metallicities \citep{tremonti04}. With our above analysis,
and scaled to solar metallicity $Z_{\sun}$, this implies that powerful
AGN feedback expels up to $0.6-14\times 10^{7} Z/{Z_{\sun}}$ \msun\
of metals per M$^{\star}$ galaxy.  Early-type nearby galaxies about a
magnitude brighter than ${\cal L}^{\star}$ have total baryonic masses
of $\sim4\times 10^{11}$ \msun, and super-solar metallicities by $\sim
0.3$ dex \citep{tremonti04}, which implies a total metal content of $\sim
10^{10}$ \msun. This indicates that MRC1138-262-like feedback could eject
a substantial fraction of the metals contained in the stars of an ${\cal
L}^{\star}$ elliptical in the local Universe.

This metal-rich outflow may contribute significantly to the overall
metallicity density at high redshift.  \citet{bouche05a} estimate
the co-moving metallicity density at z$\sim 2$ from integrating
the star-formation rate density, $\sim 4\times10^6$ \msun\
Mpc$^{-3}$. If outflow rates in powerful AGN are generally similar to
MRC1138-262, then our above estimate implies a total metal ejection
of $\sim\rho_{ej,Z}=1-30\times10^{4}\ Z/Z_{\sun}$ \msun\ Mpc$^{-3}$.
\cite{bouche05b} were able to account for $\la$50\% of the total
metallicity density at z$\sim2$ in the known population of galaxies.
Thus, AGN feedback from massive galaxies may account for of-order
a few percent to about 20\% of all metals in the IGM at z$\sim$2.
Although these estimates are highly uncertain, it appears very likely
that AGN-driven outflows are a significant contributor to the overall
metal distribution in the IGM in the early universe.

\section{Summary}

We presented a study of the spatially-resolved dynamics in the optical 
emission line gas around the z$=$2.16 powerful radio galaxy 
MRC1138-262, using the near-infrared integral-field spectrograph
SPIFFI on UT1 of the VLT. The large-scale kinematics in the emission
line nebula of this galaxy are not consistent with the signatures of
large-scale gravitational motion or starburst-driven winds. Velocities
and FWHMs of $\sim 800- 2400$ \kms\ indicate a vigorous outflow with an
estimated total energy which is orders of magnitude higher than 
suggested for starburst-driven winds. Based on timescale and energy
arguments, we conclude that the observed outflow, which is a strict
lower limit to the intrinsic amount of outflowing gas, is most
plausibly driven by the AGN. This implies a relatively efficient
interaction between AGN and interstellar medium, with efficiency
$\epsilon = {\cal O}(0.1)$, mostly mediated by mechanical energy
injection through the radio jet. Radiation-driven winds do not appear
powerful enough to explain the observed velocities over the size of
the optical line emitting gas. 

For MRC1138-262 we estimate from simple energetic arguments that the
total energy needed to drive the observed outflow of optical emission
line gas is of-order few $\times$ $10^{60}$ ergs with a mass outflow rate,
$\dot{M}\gtrsim 400$ \msun\ yr$^{-1}$.  If we assume that every luminous
QSO near the peak of their co-moving density at z$\sim$2 has such an
outflow phase with similar characteristics to MRC1138-262 we find that
powerful AGN eject $\gtrsim 5 \times 10^{56}$ ergs Mpc$^{-3}$ of energy
into the IGM and $\gtrsim 0.6-14 \times 10^{6}$ \msun\ Mpc$^{-3}$ of gas,
including $\sim 1-30\times 10^{4}$ \msun\ Mpc$^{-3}$ in metals. These
energy and masses are significant compared to the the total binding
energy and mass budget of ${\cal L}^{\star}$ early-type galaxies in
the local Universe. If MRC1138-262 is indeed archetypal, then AGN winds
have the potential to be of similar, perhaps even larger, cosmological
signficance than starburst-driven winds, especially for the most massive
galaxies. They accelerate the outflowing gas to much higher velocities
than starburst-driven winds, and much of this material and energy are
likely to escape the potential of even the most massive individual
galaxy halo.  Outflows like in MRC1138-262 have the potential to
explain the observed properties of low-redshift massive galaxies, namely
their old, and narrow range of ages and low gas fractions. Moreover,
the large mass loss rates and short timescales of their outflows can
naturally explain the observed enhancement of [$\alpha$/Fe] relative
metal abundances observed in the most massive galaxies at low redshift.
AGN-driven outflows appear to be a plausible mechanism that efficiently
suppresses star-formation within a few $\times 10^8$ yrs.  They also have
high enough energies to perhaps stop the accretion flow of matter from
the IGM into their dark matter halos.  Thus, if accretion is followed
by AGN activity, then AGN feedback may be an effective mechanism for
regulating galaxy growth \citep{croton06}.

Our study comprises only one source, but we have indirect evidence for
MRC1138-262 being far from unique. Velocities of $\sim 1000$ \kms\ are not
uncommon in the extended emission line gas of powerful radio galaxies,
but are generally extracted from longslit spectra aligned with the
radio jet axis only. Simulating a longslit spectrum from the SPIFFI data
cube, we find good agreement between MRC1138-262 and these samples. The
bolometric and radio jet luminosities of MRC1138-262 are not unusual, and
the source follows the $K-z$ relationship of high-redshift radio galaxies,
if correcting for the luminous broad \ha\ line emission. We are currently
acquiring rest-frame optical, integral-field data for a larger sample
of HzRGs, to make statistically more robust predictions. Preliminary
results strongly support the present analysis, giving further evidence
that AGN feedback plays a major role in the evolution of the most massive
galaxies in the universe.

\acknowledgements 
We would like to thank the ``SPIFFI GI'' team for carrying out the
observations, and C. de Breuck, W. Van Breugel and M. Villar-Martin
for helpful discussions.

\end{document}